\documentclass{cmspaper}

\usepackage{placeins}

\begin{document}


\begin{titlepage}

   \cmsnote{2001/000}
   \date{\today}

  \title{Prospects of detecting massive isosinglet neutrino at LHC in  the
CMS detector}

  \begin{Authlist}
    S.N.~Gninenko, M.M.~Kirsanov, N.V.~Krasnikov, V.A.~Matveev
       \Instfoot{INR}{INR, Moscow, 117312, Russia}
  \end{Authlist}

  \begin{abstract}

    A possibility to search for a heavy isosinglet (sterile) neutrino 
using its decay mode $\nu_s \rightarrow l^{\pm}~+~2~jets$ in the $S$ - channel
production $pp \rightarrow W^* + X \rightarrow l^{\pm}\nu_s + X$
in the CMS experiment is studied. The only assumption about the heavy
neutrino is its nonzero mixing with $\nu_e$ or $\nu_{\mu}$.
The corresponding CMS discovery potential
expressed in terms of the heavy neutrino mass and the mixing parameter
between the heavy and light neutrino is determined.
It is shown that the heavy neutrino with a mass up to 800 $GeV$ could
be detected in CMS. We also investigate the production of the heavy
neutrino $N_l$ mixed with $\nu_e$ and/or $\nu_{\mu}$ in the
$SU_C(3) \otimes SU_L(2) \otimes SU_R(2)\otimes U(1)$
model through the reaction $pp \rightarrow W_R + X \rightarrow l^{\pm}N_l + X$
with the same heavy neutrino decay channel as above. We find that for
$M_{W_R} < 3~TeV$ it is possible to discover the heavy neutrino
with a mass up to $0.75 \cdot M_{W_R}$.

\end{abstract} 

\end{titlepage}

\setcounter{page}{2}


\section{Introduction}

 In the Standard Model of electroweak interactions \cite{SM}, neutrinos
are the only fundamental fermions which do not have a right-handed
component that transforms as an isosinglet under the $SU(2)_L$
gauge group. However, heavy $SU_C(3) \otimes SU_L(2) \otimes U(1)$ isosinglet 
neutrinos $\nu_s$ (sterile neutrinos or isosinglet neutrinos) are predicted
in many extensions of the Standard Model \cite{SM1}.
Such isosinglet neutrino can interact with W- and Z- bosons only as a 
result of its nontrivial mixing with ordinary neutrinos.

\begin{figure}[hbtp]
  \begin{center}
    \resizebox{8cm}{!}{\includegraphics{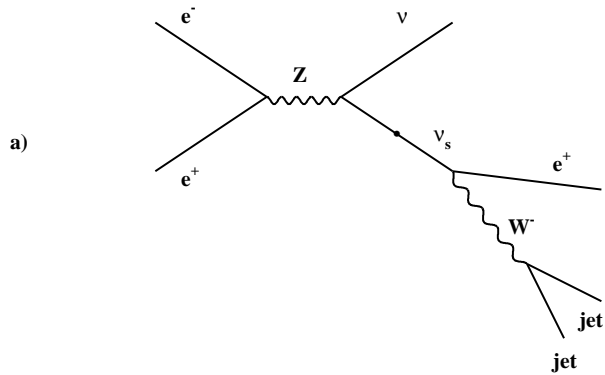}}
    \caption{Feynman diagram corresponding to the production of the isosinglet
             neutrino via $s$ - channel at LEP (the L3 search) and its
             subsequent decay.}
    \label{fig:diagrl3}
  \end{center}
\end{figure}

 An isosinglet neutrino search was performed in several
experiments \cite{oldsearch}, usually in the mass range up to a value of
the order of 1 GeV or a few GeV. The search for a heavy isosinglet neutrino
in the highest mass range (from 80 $GeV$ to 200 $GeV$) was performed by
the L3 collaboration at LEP \cite{L3} using the reaction
\begin{equation}
 e^+e^- \rightarrow Z^* \rightarrow \nu (\bar{\nu}_s 
\rightarrow e^+(W^- \rightarrow jet~ jet)).
\end{equation}
The corresponding Feynman diagram is shown in Fig.~\ref{fig:diagrl3}.
No signal was observed and the limit on the mixing strength with
$\nu_e$ was obtained as a function of the heavy neutrino mass. For a
mass $m_N = 80~GeV$ (maximal sensitivity point) the 95\% C.L. limit
on $|U_e|^2$ is $0.2 \cdot 10^{-2}$.

\begin{figure}[hbtp]
  \begin{center}
    \resizebox{8cm}{!}{\includegraphics{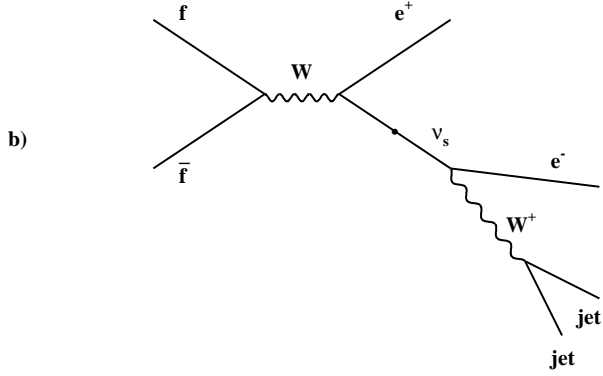}}
    \caption{Feynman diagram corresponding to the production of the isosinglet
             neutrino via $s$ - channel at LHC (the CMS search) and its
             subsequent decay.}
    \label{fig:diagrl4}
  \end{center}
\end{figure}

 In this paper we study the possibility to detect heavy isosinglet neutrino
at the LHC in the CMS detector. In our study we consider the case when
the heavy neutrino mixes mainly with $\nu_e$ or $\nu_{\mu}$.
As a result of the mixing of the heavy sterile neutrino with ordinary neutrinos 
the former can be produced at the LHC in the reaction 
\begin{equation}
pp  \rightarrow l^{\mp} (W^{*+} \rightarrow \nu_s l^{+}) ~+~X
\end{equation}
The heavy neutrino decay mode
\begin{equation}
\nu_s \rightarrow l^{\pm}W^{\mp} \rightarrow l^{\pm} + 2~jets
\end{equation}
is the most interesting for the sterile neutrino detection at the LHC. 
This decay mode leads to the signature with 2 isolated leptons and 2 jets.
The invariant mass of the $l^{\pm} ~+2~jets$ distribution $m_{inv}(l^{\pm}
~+2~jets)$ has a resonance structure due to the assumed existence 
of the heavy neutrino decay mode $\nu_s \rightarrow~l^{\pm}~ + ~2~jets$ that 
allows to separate the signal from the background. The relevant Feynman
diagram is shown in Fig.~\ref{fig:diagrl4}. It corresponds to the charged
current decay of a heavy neutrino. $W$ boson in the diagram is real. There is a
similar diagram with two muons in the final state. Our results for
a dominant mixing with $\nu_e$ practically coincide with the corresponding
results for a dominant mixing with $\nu_{\mu}$.
Two cases are possible: Dirac heavy neutrino conserving lepton number
(the diagram shown in Fig.~\ref{fig:diagrl4} corresponds to this case)
and Majorana heavy neutrino
not conserving the lepton number. In the latter case the charged leptons in the
diagram Fig.~\ref{fig:diagrl4} can have the same sign.
We study here both cases. Note that the sensitivity for the neutral current
decay mode of the heavy neutrino is significantly worse due to small
branching ratio of $Z$ boson decay into leptons.
The CMS isosinglet neutrino discovery potential depends both on the mixing 
parameter of the isosinglet neutrino with ordinary neutrinos and isosinglet 
neutrino mass. We find that it would be possible to discover the isosinglet 
neutrino with a mass up to 800 $GeV$. In out estimates we took the total
luminosity $L_t = 30~fb^{-1}$  (the first 2-3 years of the LHC operation).

 The possibility of detection at LHC of a heavy right-handed neutrino $N_l$
considered in the framework of the left-right symmetric model
$SU_C(3) \otimes SU_L(2) \otimes SU_R(2)\otimes U(1)$ \cite{model1} has
been discussed in ref. \cite{model2} (see also related papers \cite{model3}).
The corresponding production reactions could be
\begin{equation}
 pp \rightarrow W_R + X \rightarrow l^+N_l + X \rightarrow
                                           l^+l^+q \bar q^{'} + X
\end{equation}
\begin{equation}
 pp \rightarrow Z_R + X \rightarrow N_l N_l + X  
\rightarrow l^+l^+q \bar q^{'}q^{''} \bar q^{'''} + X
\end{equation}

 We studied the production and detection in CMS of such
$SU_C(3) \otimes SU_L(2) \otimes SU_R(2)\otimes U(1)$ heavy neutrino
mixed with ordinary neutrinos. The production reaction is
$pp \rightarrow W_R + X \rightarrow l^{\pm}N_l + X$, the heavy neutrino decay
channel being the same as for $\nu_s$. The CMS discovery potential in
this case is expressed in terms of the $M_{W_R} - M_{N_l}$ region, in
which the discovery is possible.

\section{The simulation of signal events.}

\begin{figure}[hbtp]
  \begin{center}
    \resizebox{15cm}{!}{\includegraphics{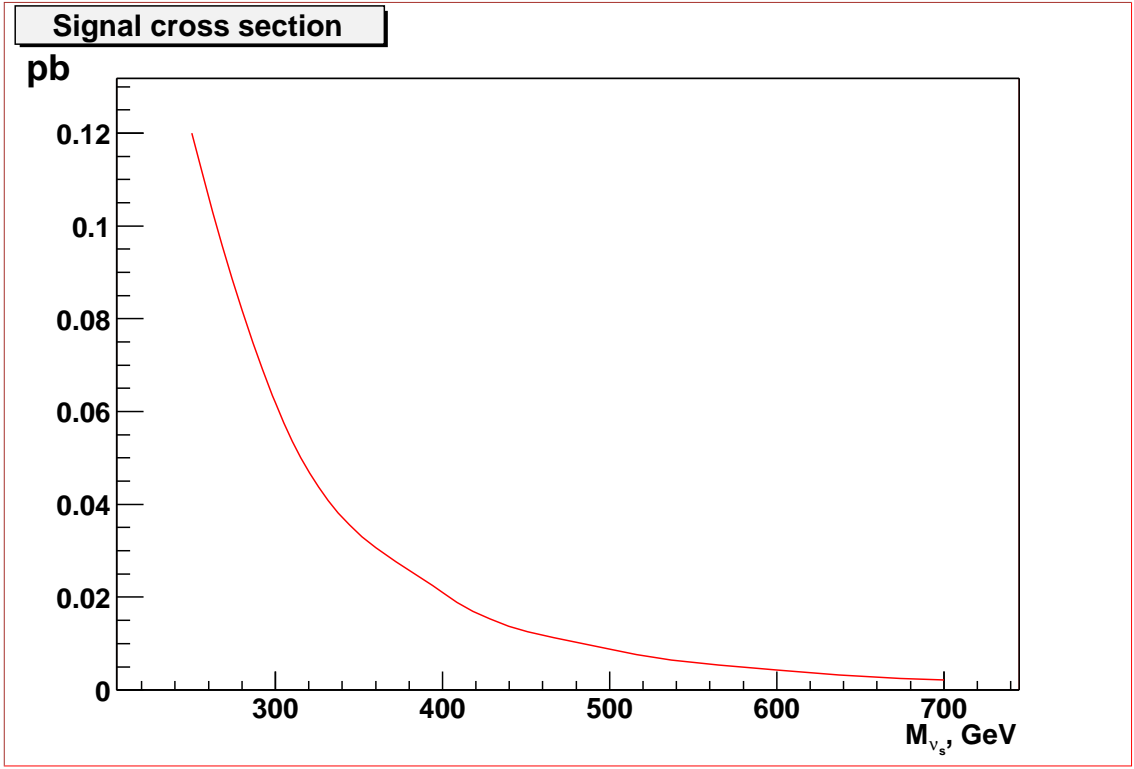}}
    \caption{Cross section of heavy neutrino production multiplied by the
             branching ratio of its subsequent decay into a charged
             lepton and two jets for the maximal mixing 
             $|U_l|^2 =1$}
    \label{fig:cs}
  \end{center}
\end{figure}

 In our simulations we used PYTHIA 6.152 \cite{pythia}, modified for the
simulation of signal events. The dependence of the signal cross section
on the heavy neutrino mass is shown in Fig.~\ref{fig:cs}. 

 The CMS fast detector simulation program CMSJET 4.703 \cite{cmsjet} was used.
The following cuts on the transverse momentum were applied: 20 $GeV$ for
electrons and muons, 40 $GeV$ for jets. Lower cut for muons, which is
default in CMSJET, does not improve the sensitivity. The isolation of
leptons in the calorimeter was determined using the cone radius 0.3 and
allowing maximal $E_t$ of 5 $GeV$ in the cone (the CMSJET default).

\section{The selection criteria and candidate event variables.}

\begin{figure}[hbtp]
  \begin{center}
    \resizebox{15cm}{!}{\includegraphics{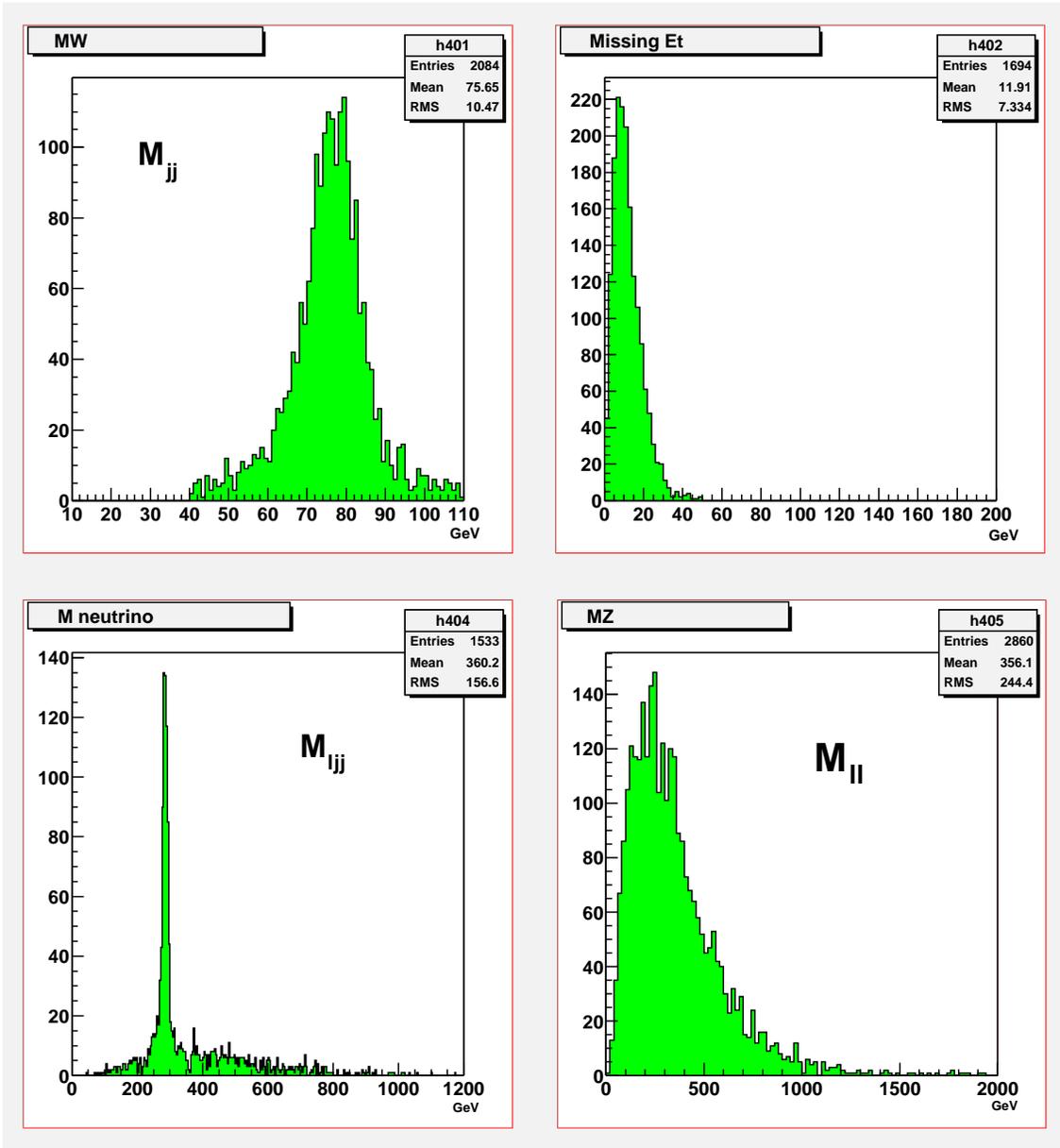}}
    \caption{Distributions of signal events with a heavy neutrino
             mass $m_{\nu_s}=300~GeV$ (arbitrary normalization)}
    \label{fig:sign300}
  \end{center}
\end{figure}

 In our analysis we proceeded through the following steps:
\begin{itemize}
\item Events with 2 isolated leptons of the same flavour and
      opposite signs were selected. Events with more than 2 leptons
      were rejected. The invariant mass $M_{ll}$ of these two leptons
      is the first candidate event variable.

\item Events with at least 2 jets were selected. From all jets a pair
      with invariant mass $M_{jj}$ closest to the mass of the W boson
      was chosen. $\Delta M_W = |(M_{jj} - m_W)|$ is the second variable.

\item From the 4-momenta of these two jets and the 4-momentum of a lepton
      the invariant mass $M_{\nu}^{cand}$ is calculated. A peak in the
      distribution of this mass is to be searched for (Fig.~\ref{fig:sign300}).

\item The  $E_t^{miss}$ of an event is the last variable of our analysis.

\end{itemize}

\section{The background.}

 The $ZW$ production is the obvious source of background events. They were
simulated with standard PYTHIA with lepton decay modes of $W$ and hadron
decay modes of $Z$ forbidden. The variable $M_{ll}$ was used to suppress
this kind of background. In ref \cite{L3} events with $M_{ll}$ close to
the $Z$ mass central value were rejected. However, the tail of
the $Z$ mass distribution is rather long. At the same time the
signal events usually have big $M_{ll}$ (Fig.~\ref{fig:sign300}). For this
reason there was simply a lower cut on $M_{ll}$ at values well above the
$Z$ mass central value.

\begin{figure}[h]
\begin{center}
    \resizebox{15cm}{!}{\includegraphics{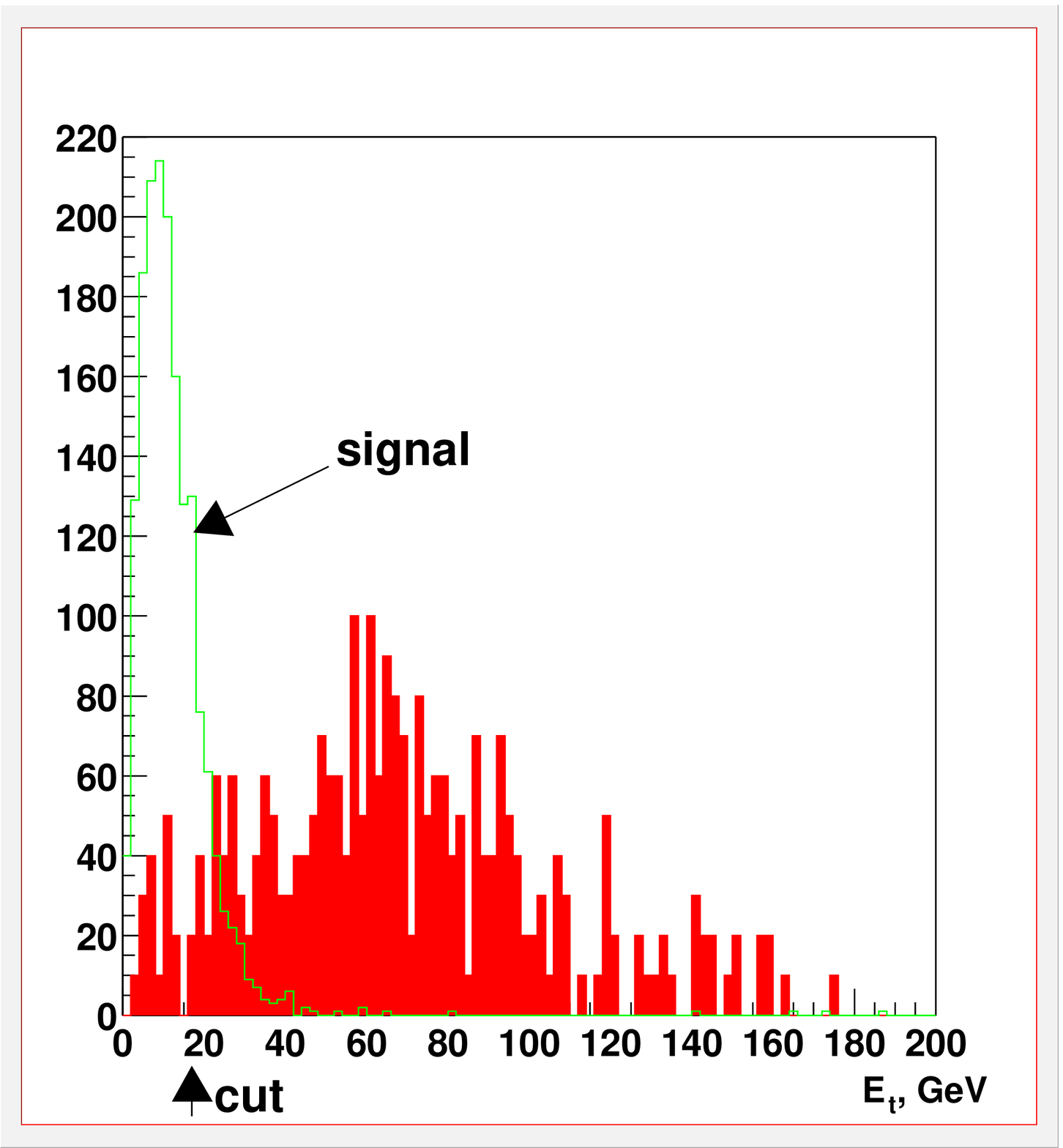}}
     \caption{The use of the $E_t$ cut for the suppression of the
              $t\bar t$ background}
      \label{fig:etcut}
   \end{center}
\end{figure}

 The $t \bar t$ production turned out to be one of the most dangerous
backgrounds. The first estimates were made with PYTHIA, but the final
ones, used in the sensitivity estimations, with TOPREX \cite{Slab}.
This program correctly takes into account the spin correlations between
the $t$ and $\bar t$ and uses TAUOLA code for $\tau$ decays. TOPREX gives
$\approx 15\%$ smaller number of initial (with loose cuts) candidate
events than PYTHIA. Only leptonic $W$ decay modes were allowed (including
$\tau \nu_{\tau}$). It was checked that other decay modes do not contribute.
One of the most powerful cuts for the rejection of this background is the
upper $E_t^{miss}$ cut (Fig.~\ref{fig:etcut}).

\begin{figure}[h]
\begin{center}
    \resizebox{15cm}{!}{\includegraphics{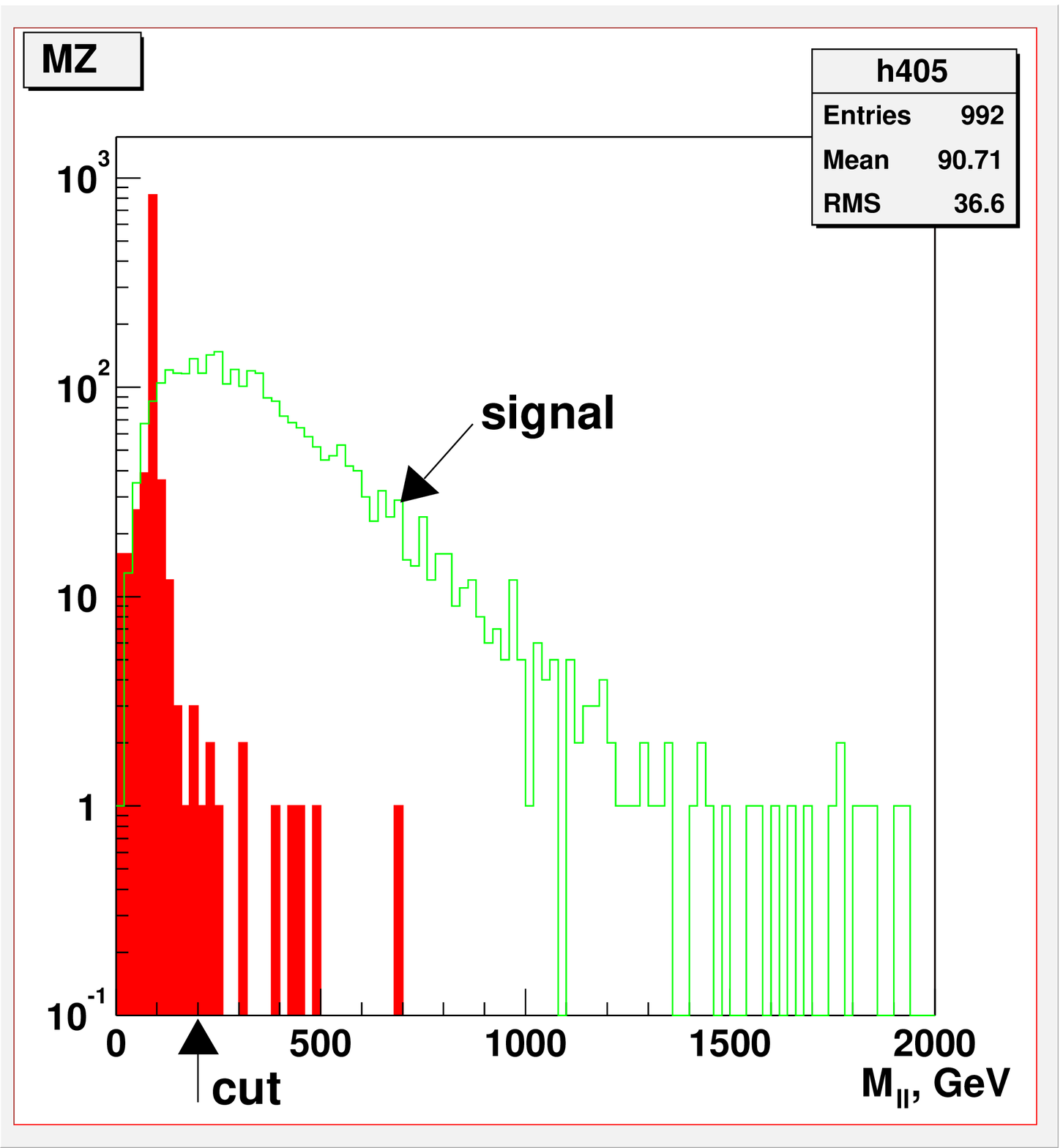}}
     \caption{The use of the $M_{ll}$ cut for the suppression of the
              $Zg$ background}
      \label{fig:mzcut}
   \end{center}
\end{figure}

 Another dangerous background is the $Z + jet$ production. This process
has a large cross section and requires a lot of simulation with PYTHIA.
In order to reduce the required CPU time only events with $Q_t > 20~GeV$
were simulated. It was checked with loose cuts that the above cut does
not change the estimated number of background events for a given
luminosity. This background is suppressed by the same cut as the
$ZW$ production and by the $P_t$ cuts on leptons and jets.
In Fig.~\ref{fig:mzcut} one can see the long tail of the $Z$ invariant
mass in this background process and how the $M_{ll}$ is used for its
suppression.

 The other possible sources of background are the $ZH$ and $WH$ productions.
In our calculations we took $m_H = 150~GeV$. 
However, the cross
sections are small and this background is not dangerous.

\begin{figure}[hbtp]
  \begin{center}
    \resizebox{15cm}{!}{\includegraphics{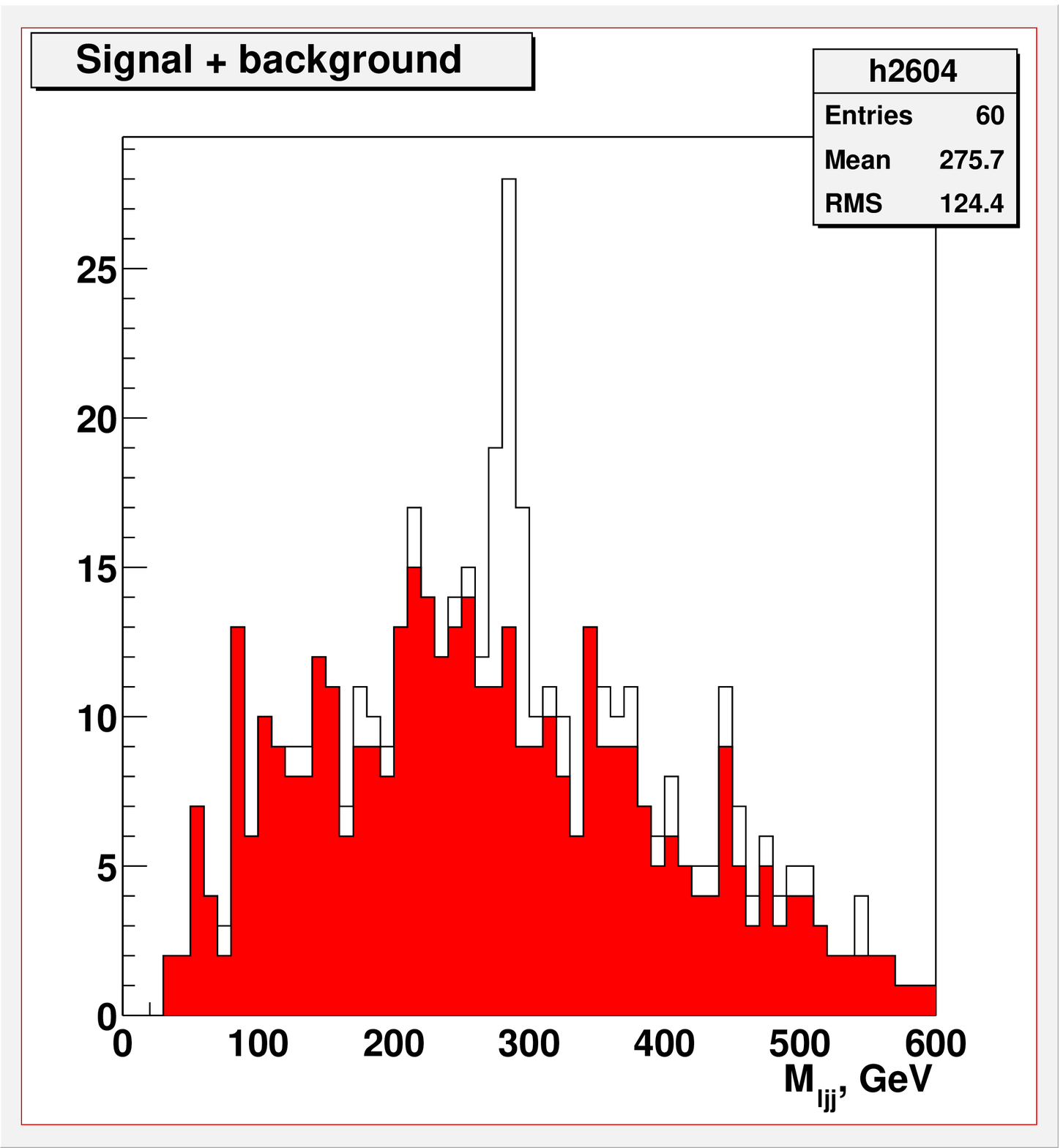}}
    \caption{Invariant mass distribution of candidate events.
             The heavy neutrino with a mass of 300 $GeV$ is mixed with
             $|U_l|^2=0.1$ with electron or muon neutrino. Shaded
             region - only background events. The normalization
             corresponds to $L_t=30fb^-1$}
    \label{fig:figcms}
  \end{center}
\end{figure}

In Fig.~\ref{fig:figcms} the distribution of candidate event invariant masses
expected in CMS with $30 fb^{-1}$ is shown, assuming a 300 $GeV$ heavy
neutrino mixed with $|U_l|^2=0.1$ with electron or muon neutrino.

 The only found sources of events with same sign
leptons that could be a background to the production of heavy Majorana
neutrino are the $ZH$ and $WH$ productions. Due to small cross section
of these processes the search for Majorana neutrino is almost a
background-free search.

\section{The sensitivity estimation.}

 For each value of $M_{\nu}$ a probability $\epsilon$ of signal events to pass
all cuts and to have $M_{\nu}^{cand}$ in some range around $M_{\nu}$
("good events") was calculated. The total number of background events
passing the same cuts and having $M_{\nu}^{cand}$ in the same range was
calculated for a given luminosity. 
The sensitivity and the discovery potential of the CMS 
experiment to the search for 
heavy isosinglet neutrino has been estimated using the method of 
ref. \cite{sensbk}.  Using the efficiency
$\epsilon$ the sensitivity in terms of cross section was calculated.
This cross section divided by the cross section for the maximal mixing
strength $|U_l|^2=1$ gives a limit in terms of a mixing strength.
For each value of $M_{\nu}$ the cuts and the $M_{\nu}^{cand}$ range
were optimized to obtain the best sensitivity.

\begin{figure}[hbtp]
  \begin{center}
    \resizebox{15cm}{!}{\includegraphics{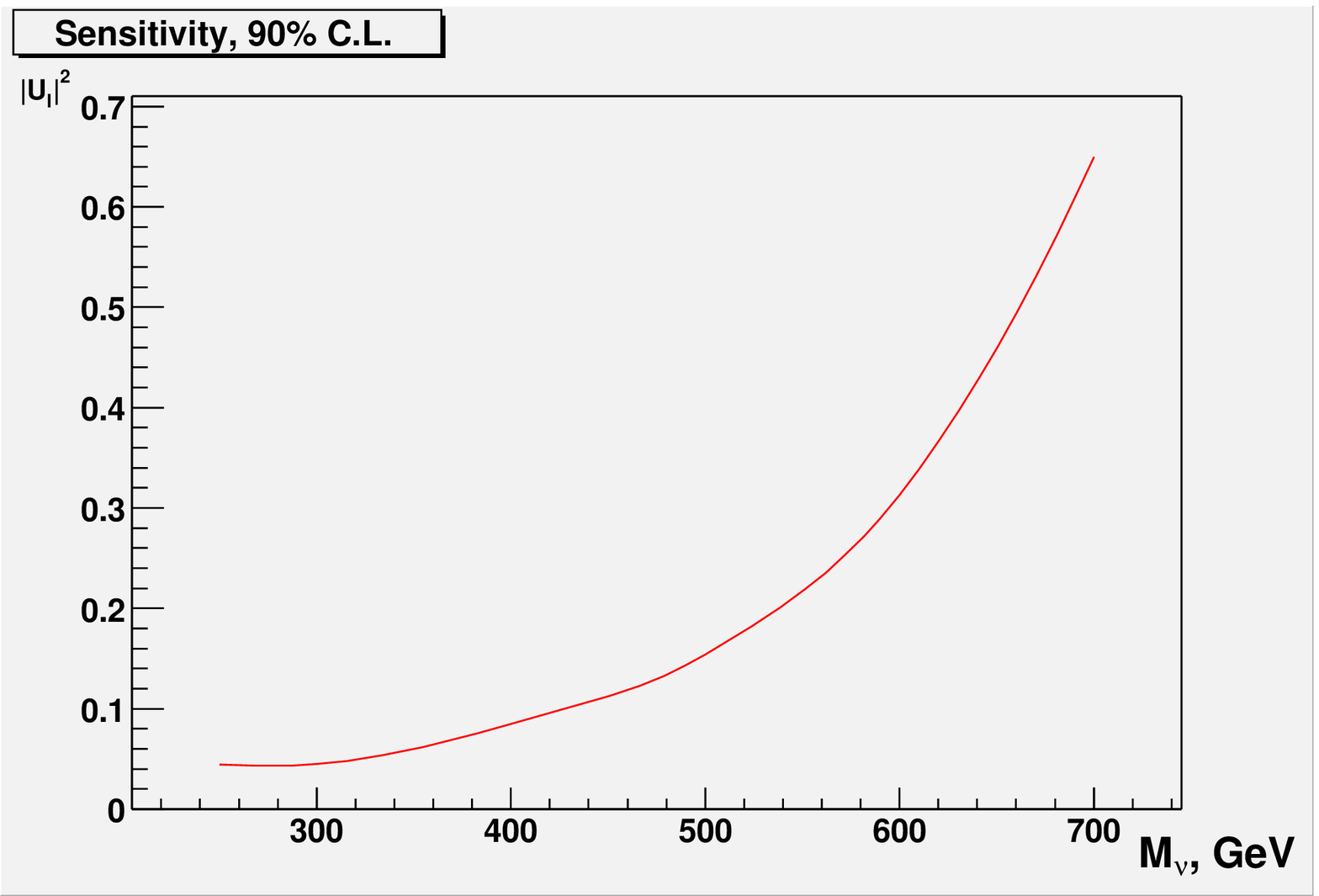}}
    \caption{CMS 90\% C.L. sensitivity for opposite sign lepton pairs.}
    \label{fig:sensit}
  \end{center}
\end{figure}

 The 95\% C.L. sensitivity as a function of $M_{\nu}$ is shown in
Fig.~\ref{fig:sensit}. Within 10\% it is the same for both the cases of
mixing with an electron and muon neutrino.

\begin{figure}[hbtp]
  \begin{center}
    \resizebox{15cm}{!}{\includegraphics{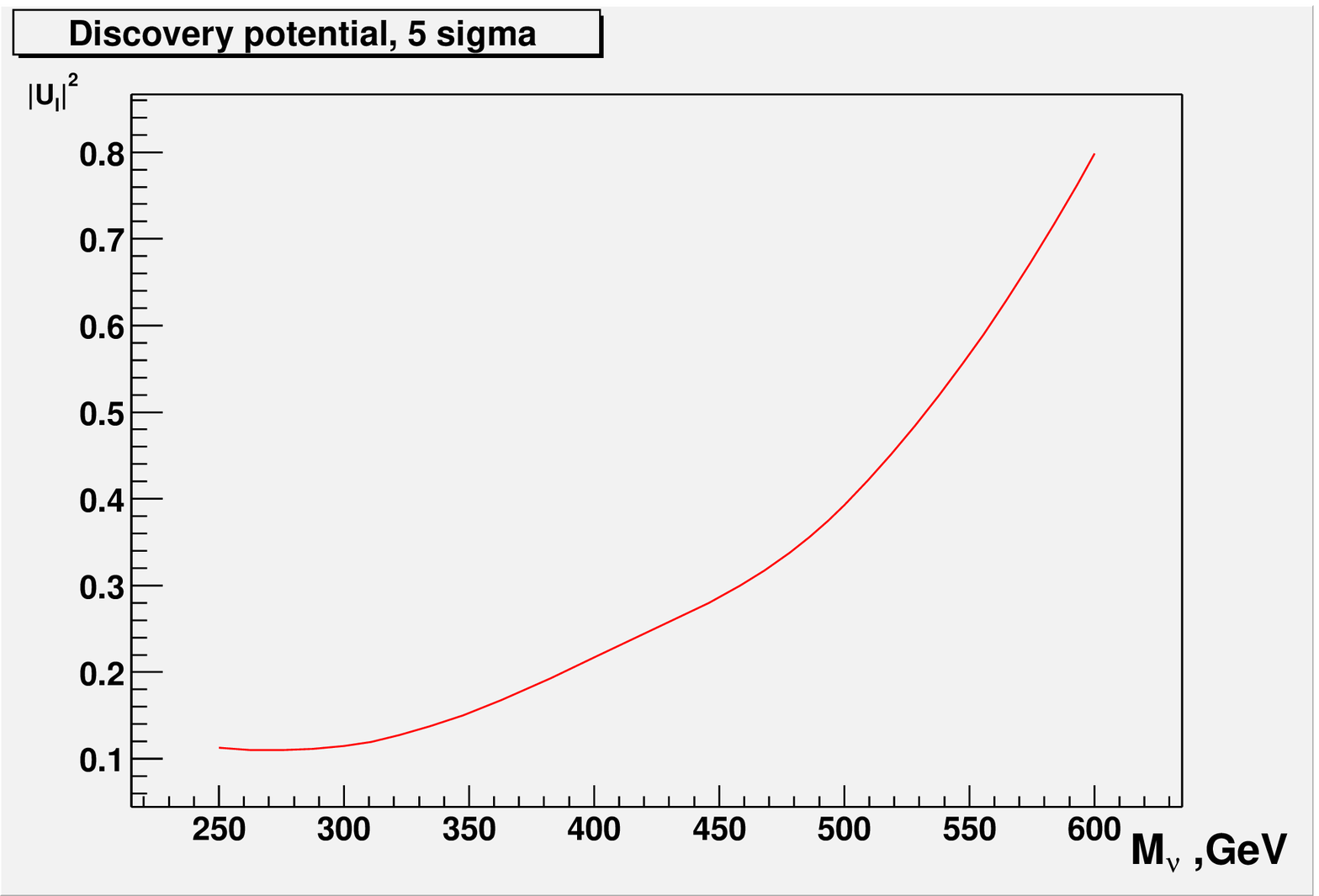}}
    \caption{CMS $5\sigma$ discovery plot for opposite sign lepton pairs.}
    \label{fig:sensitbk}
  \end{center}
\end{figure}

 The CMS $5\sigma$ discovery potential estimated by the same method
is shown in Fig.~\ref{fig:sensitbk}.

\begin{figure}[hbtp]
  \begin{center}
    \resizebox{15cm}{!}{\includegraphics{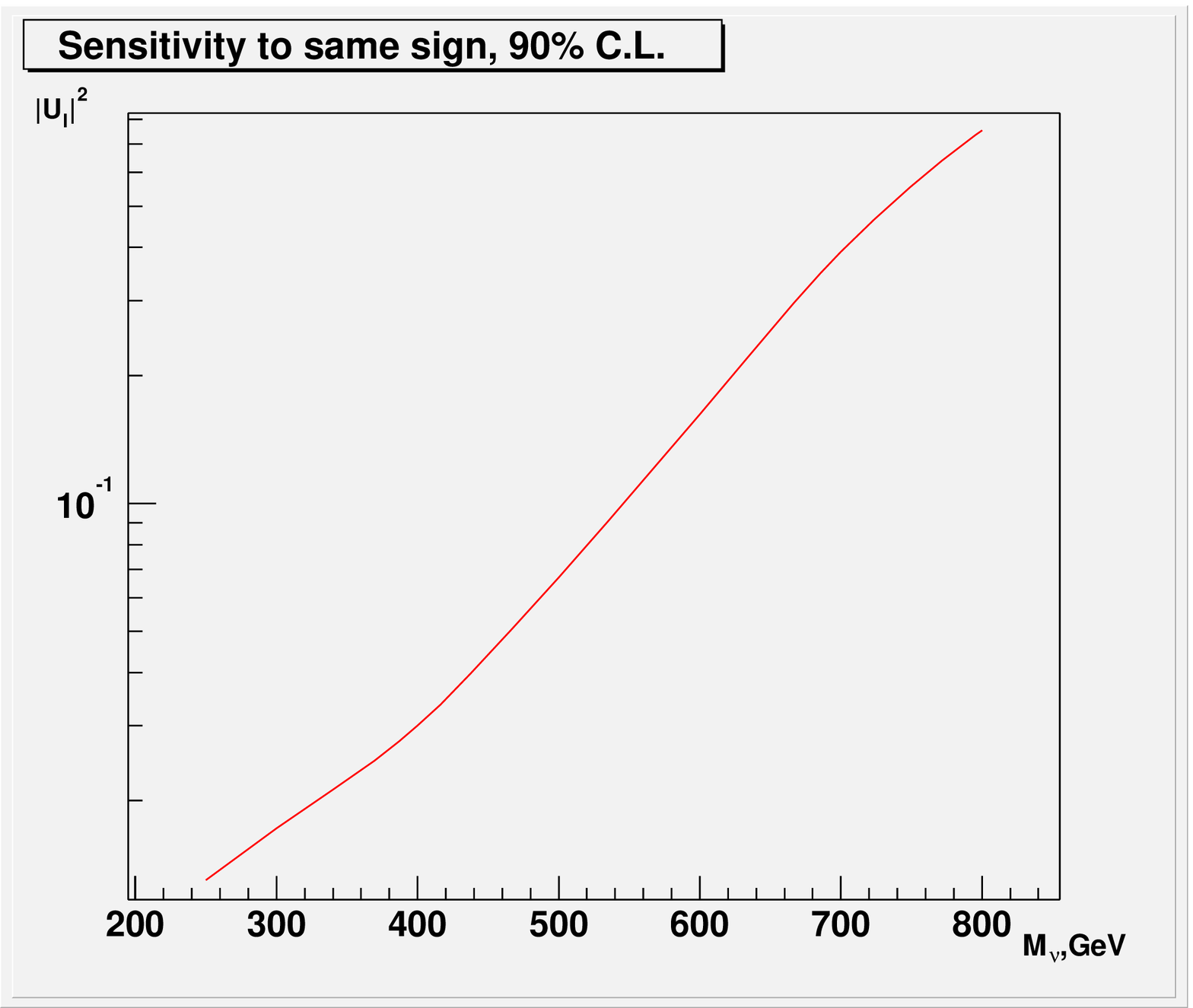}}
    \caption{CMS 90\% C.L. sensitivity for same sign lepton pairs}
    \label{fig:sensitss}
  \end{center}
\end{figure}

\begin{figure}[hbtp]
  \begin{center}
    \resizebox{15cm}{!}{\includegraphics{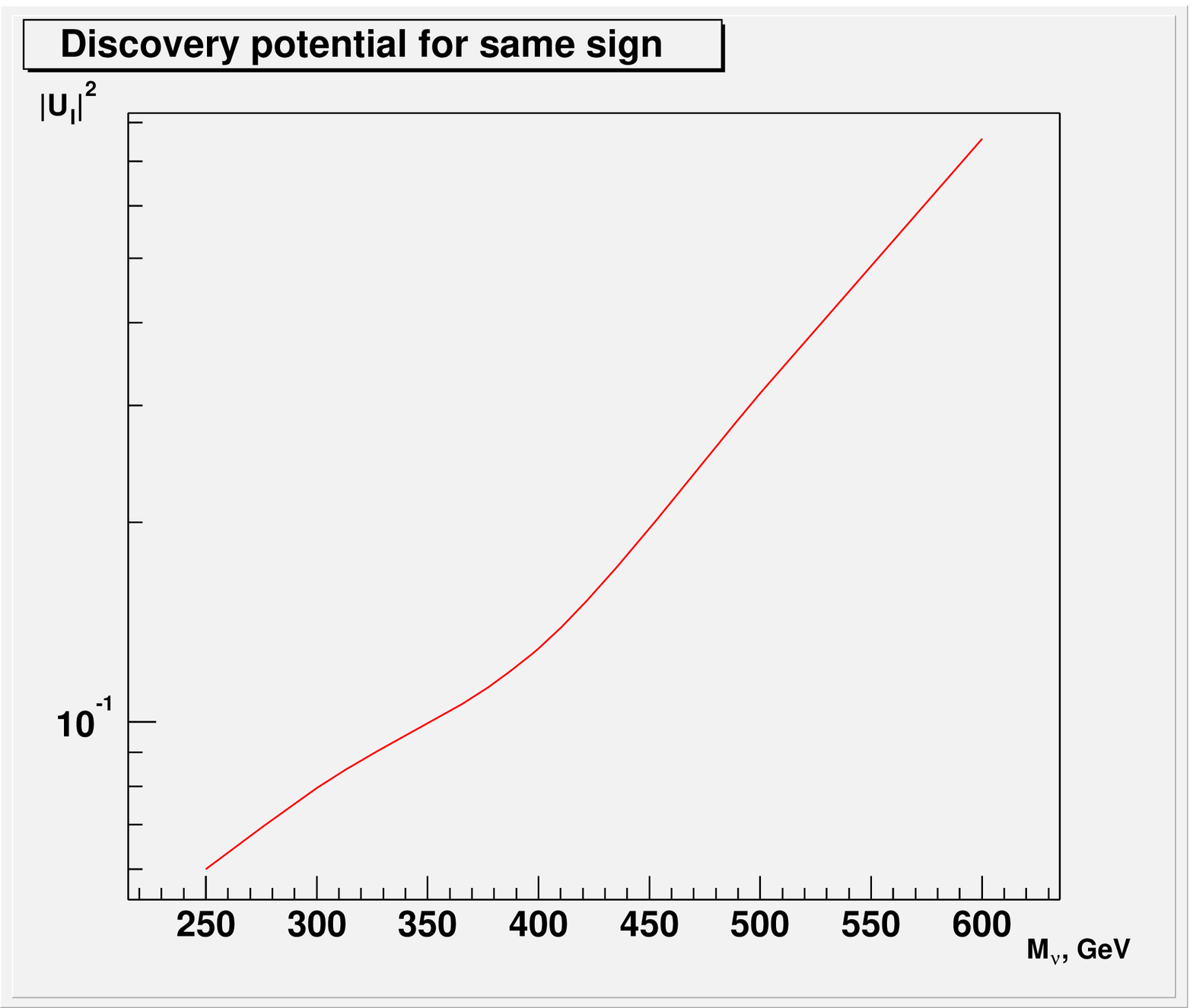}}
    \caption{CMS $5\sigma$ discovery potential for same sign lepton pairs}
    \label{fig:sensitss5s}
  \end{center}
\end{figure}

 The CMS sensitivity to the heavy neutrino production violating lepton
number (Majorana case) is shown in Fig.~\ref{fig:sensitss} and the
CMS discovery potential in Fig.~\ref{fig:sensitss5s}.
Here we assumed that 50\% of heavy neutrinos decay with lepton
number violation that leads to same sign lepton pairs.
It is almost zero background search. Some background events were found in
the $WH$ sample, but their contribution is very small due to low cross
section. Events with our signature and same sign leptons can be produced
in the chain decays of $t\bar t$, one lepton being produced at the first
step, in the $t$ or $\bar t$ decay, and another at the second step,
in a B meson decay (here $B^0$ oscillations should be taken into account).
However, this background is strongly suppressed by the requirement
of isolation in combination with a lepton $P_t$ cut and by the cut on
the missing $E_t$.

\section{The heavy neutrino production in the
         $SU_C(3) \otimes SU_L(2) \otimes SU_R(2)\otimes U(1)$ model.}

\begin{figure}[h]
\begin{center}
    \resizebox{15cm}{!}{\includegraphics{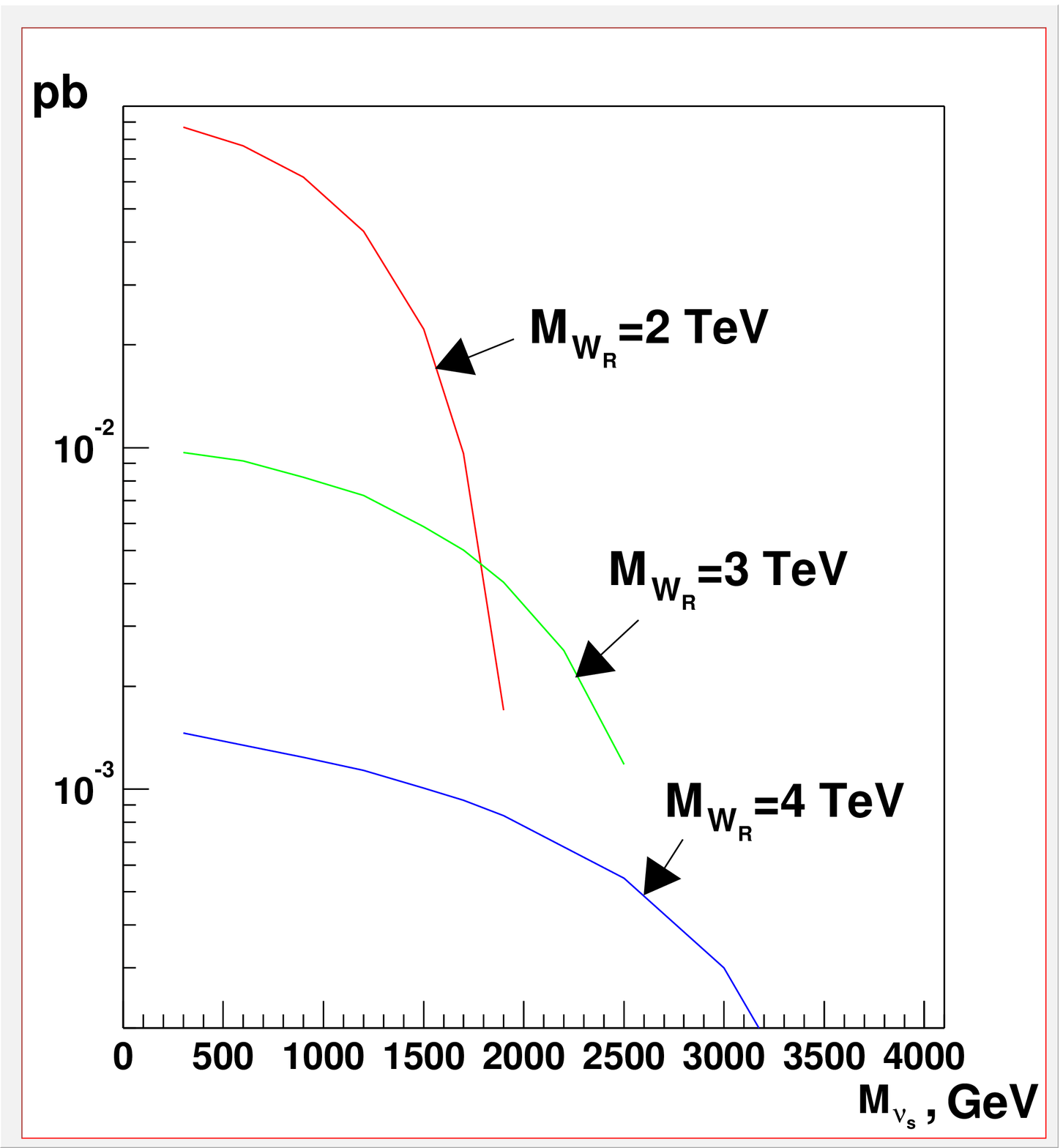}}
     \caption{The dependence of the value
              $\sigma(pp \rightarrow W_R \rightarrow l^{\pm}N_l)
              \cdot Br(N_l \rightarrow l^{\pm}~+~2~jets)$
              on the heavy neutrino mass}
      \label{fig:cswr}
   \end{center}
\end{figure}

 Another scenario with a heavy neutrino was proposed in
\cite{model2}. In this scenario the heavy neutrino $N_l$ can be produced
through its coupling with
the heavy $W_R$ boson, the latter being coupled with quarks and three heavy
neutrinos $N_l$. One of the $N_l$ neutrinos is relatively light
(a few hundred $GeV$) and can decay through
its mixing with $\nu_e$ or $\nu_{\mu}$, other $N_l$ are very heavy.
We assumed that the lightest $N_l$ is $N_e$. We checked that the
results with $N_{\mu}$ being the lightest do not differ much.  
In this scenario we have to look for a heavy Majorana neutrino producing same
sign leptons. In Fig.~\ref{fig:cswr} the cross sections for different
masses of $W_R$ are shown. They don't depend on the heavy neutrino
mixing parameters. We don't study $W_R$ masses below 1500 $GeV$,
assuming that they are excluded by indirect analyses \cite{model2}.

 In ref. \cite{model2} the $t\bar t$ production was found to be the most
dangerous background. The level of this background was plotted as a function
of the lepton $P_t$ cut and was not yet zero with a 20 $GeV$ cut. However,
in that work the missing $E_t$ cut was not used. In our study this cut
kills the $t\bar t$ events survived after lepton isolation cuts and the
lepton $P_t$ cut.

\begin{figure}[h]
\begin{center}
    \resizebox{15cm}{!}{\includegraphics{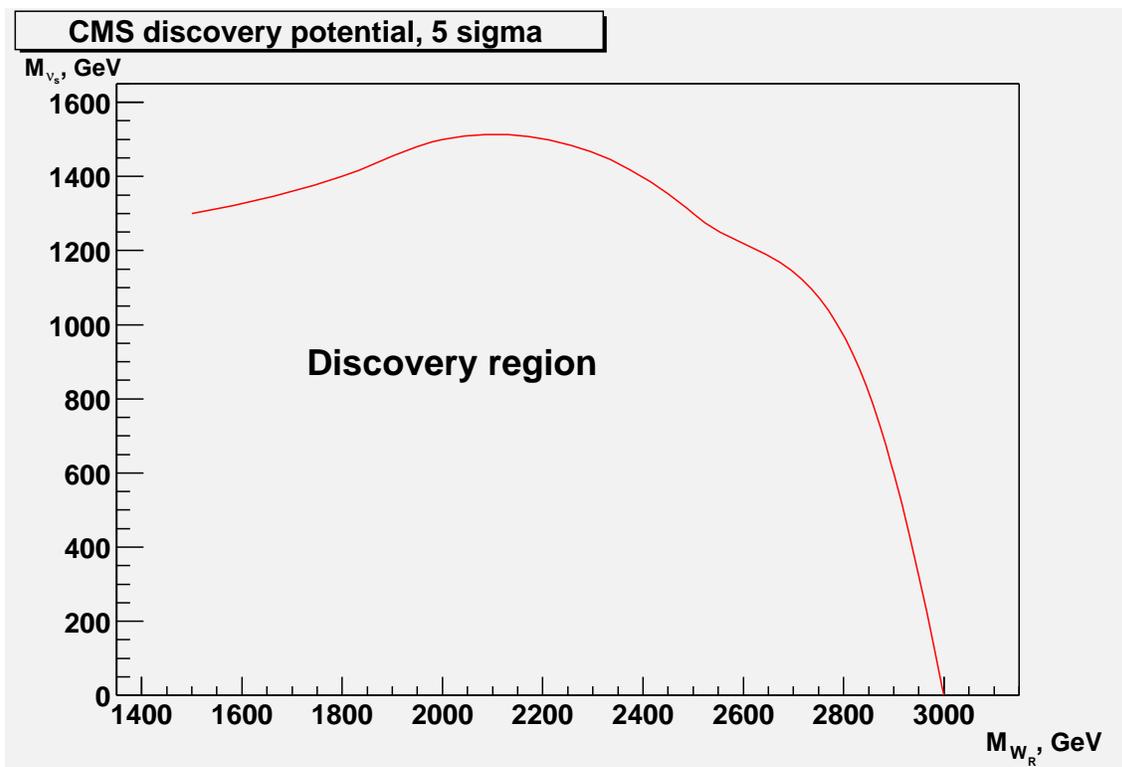}}
     \caption{CMS discovery potential for the left-right symmetric model
              and $L_t = 30~fb^{-1}$}
      \label{fig:reswr}
   \end{center}
\end{figure}

\section{Conclusion.}

 In this paper we presented the results of our study of a possibility
to detect the heavy isosinglet neutrino $\nu_s$ at LHC in the CMS detector. 
In the first part of our study we assumed only the nonzero mixing of this
neutrino with $\nu_e$ or $\nu_{\mu}$ neutrinos. We found that it
is possible to detect sterile neutrinos with masses up to 800 $GeV$
(for the masses near this value and above it a mixing $|U|^2$
sufficient for the detection becomes unrealistically high).
 In the second part we assumed the existence of additional interaction
with a participation of a heavy neutrino, namely the
$SU_C(3) \otimes SU_L(2) \otimes SU_R(2) \otimes U(1)$ model with
a heavy $W_R$ boson \cite{model2}. We assumed also that this heavy
neutrino $N_l$ is mixed with ordinary neutrinos $\nu_e$ or $\nu_{\mu}$,
that makes possible the same decay mode as for $\nu_s$.
We found that for $M_{W_R} < 3~TeV$ it is possible to discover
in CMS the heavy neutrino $N_l$ (l is electron or muon) with a mass
up to 1500 $GeV$. 

\FloatBarrier

\section{Acknowledgments}

 We are grateful to S. Slabospitsky for his help in the work with the
program TOPREX and for useful discussions.
 We are indebted to the participants of Daniel Denegri seminar on physics
simulation at LHC for useful comments.

\end{document}